\documentclass[fleqn]{mnras}
\usepackage{natbib}
\usepackage{ifpdf} 

\usepackage[T1]{fontenc}

\usepackage{graphicx}	
\usepackage{amsmath}	
\usepackage{amssymb}	
\usepackage{mathptmx}
\usepackage{multicol}
\usepackage{txfonts}

\usepackage{subcaption}

\title[Additional EBL at Optical-NIR band]{Prospects of additional contribution at Optical-NIR band of EBL in the light of VHE spectra}

\author[N. Mankuzhiyil et al.]{
Nijil Mankuzhiyil$^{1}$\thanks{E-mail: mankuzhiyil.nijil@gmail.com}, Massimo Persic$^{2,3}$,  Alberto Franceschini$^{4}$ \\
$^{1}$ Astrophysical Sciences Division, Bhabha Atomic Research Centre, Mumbai 400085, India \\
$^{2}$ INAF Osservatorio Astronomico di Padova, Vicolo dell'Osservatorio 5, I-35122, Padova, Italy\\
$^{3}$ INFN-Trieste, via A.Valerio 2, I-34127 Trieste, Italy\\
$^{4}$ University of Padova, Physics and Astronomy Department, Vicolo Osservatorio 3, I-35122 Padova, Italy
}

\date{Accepted XXX. Received YYY; in original form ZZZ}

\pubyear{2022}

\begin{document}
\label{firstpage}
\pagerange{\pageref{firstpage}--\pageref{lastpage}}
\maketitle

\begin{abstract}

The Extragalactic Background Light (EBL) that spans the UV-IR band originates from direct and dust-reprocessed starlight 
integrated over the history of the Universe. EBL measurements are very challenging due 
to foreground emission like  the zodiacal light { and interplanetary dust emission. Indeed, some optical/NIR direct measurements overpredict EBL models 
based on galaxy counts.} On the other hand, there is  some debate on possible additional components of the Optical-NIR photon 
density: e.g., population-III stars, axion-photon decay, direct collapse of black holes, intra-halo light etc. Owing to the 
absorption of Very High Energy (VHE) $\gamma$ rays by interaction with EBL photons, we study the prospects of accommodating an 
additional population of EBL sources in the Optical-NIR band { on top of the standard galaxy-count--based component}. To this aim we use 105 VHE spectra of 37 blazars with known redshifts, $0.03<z<0.94$. We correct the observed spectra for absorption by our model EBL. By requiring the intrinsic spectra to be non-concave and with a VHE spectral index $>$1.5, we estimate, at different wavelengths, upper limits to the additional low energy photon fields which would contribute to the absorption of $\gamma$-rays. Considering these limits we suggest that there is room for photons from Pop III stars and axion-like particle (ALP) annihilation. However, these additional hypothetical photon fields are bound to fall significantly below direct published EBL measurements by several instruments, and therefore our limits are either in tension or even inconsistent with such measurements.

\end{abstract}




\section{Introduction}

The Extragalactic Background Light (EBL), that spans the UV-IR band, is the integrated diffuse starlight generated throughout the evolution 
of the Universe. Its spectral energy distribution (SED) { consists} of two humps: one arising from direct starlight that peaks at $\lambda \sim 1\,\mu$m (optical background), the other arising from dust-driven absorption of the UV starlight and reemission peaking at $\lambda 
\sim 100\,\mu$m (infrared background).
%
Direct measurements of the EBL are hindered by foreground light, mostly the zodiacal light. Discrete sources like stars (primarily at $\sim$ 
1.25 to 3.5 $\mu$m), scattered emission from interplanetary dust (at $\sim$1.25 to 140$\mu$m), and interstellar dust (at $\gtrsim$60 $\mu$m) 
are also relevant sources of foreground light \cite{back}. Consequently, EBL measurements carry significant uncertainties  \citep{Matsuoka, 
Mattila} and indirect methods are banked upon in order to deduce the EBL photon density. 

Galaxy evolution models may be used to determine the EBL SED at different redshifts. These models can be primarily classified as 
{\it forward evolution} and {\it backward evolution} models. The former evolve galaxies in time from cosmological initial conditions 
using semi-analytical models of galaxy formation, stellar evolution, and dust-driven absorption/remission of photons [eg: \cite*{Finke2010, 
Gilmore2012}]. The latter integrate the observed galaxy counts, which provide model-independent lower limits to the EBL, and are 
extrapolated to higher redshifts using number counts, redshift distributions, star formation rates, luminosity functions, or directly 
observed galaxy populations at different wavelengths [eg: \cite*{fr2008, fr2017, dom2011}].

On the other hand, even before extragalactic VHE $\gamma$-ray photons were detected an indirect method relying on Very High Energy (VHE; 
E $\gtrsim$ 30\,GeV) observations of blazars was proposed to measure the EBL density \citep{firstgamma}. This method relies on the concept 
that the interaction of VHE photons emitted by faraway sources with the intervening EBL leads to the formation of electron-positron pairs, 
$\gamma \gamma \rightarrow e^-e^+$. This results in an energy-dependent attenuation of the measured VHE flux. The first detections of VHE 
photons from Mrk\,421 and Mrk\,501 (by the Whipple and HEGRA telescopes respectively) yielded upper limits on the EBL density \citep{mrk421, 
mrk501}. Detections of relatively high-$z$ blazars like H\,2356-309 ($z=0.165$), 1ES\,1101-232 ($z=0.186$) by HESS, 1ES\,1011+496 ($z=0.212$) 
by MAGIC, and PKS\,1441+25 ($z=0.939$) by VERITAS yielded plausible single-VHE-spectrum--based upper limit estimations on the EBL \citep{2356, 
1011,1441}. Modeling the optical-to-GeV SED of blazars, and comparing actual VHE data with the models extrapolated into the VHE range (where 
EBL absorption is effective), provides a way to estimate the  actual EBL density at the source redshift \citep{nijil}. 

A larger set of blazar VHE spectra can lead to more robust estimates of the EBL. \cite{mazinebl} considered 13 observed VHE spectra to derive 
upper limits in the range 0.4--100\,$\mu$m. Later \cite{meyerebl} followed a similar approach using a sample of 23 VHE spectra together with 
the corresponding spectral parameters obtained from Fermi-LAT data. \cite{biteauwilliams}, using 20 years of $\gamma$-ray observations,  
investigated the EBL and possible anomalies. Deducing 17 VHE spectra (H.E.S.S. data) of 7 blazars, \cite{hess2013} evaluated a normalization 
factor $\sim$1.3 for the EBL optical depths of the \cite{fr2008} model. Similarly, using 32 VHE spectra (MAGIC data) of 12 blazars 
\cite{abelardo} estimated EBL optical depths, which turned out to be comparable, within factors of 0.82-1.23, to those of several EBL models. 
With a different approach, \cite{hess2017} employed 19 VHE blazar spectra (H.E.S.S. data) with redshift z$<$0.29 to estimate the EBL SED 
independent of existing EBL models.

\section{Ambiguity in the optical-NIR peak}

\begin{figure}
   \includegraphics[width=1.1\columnwidth]{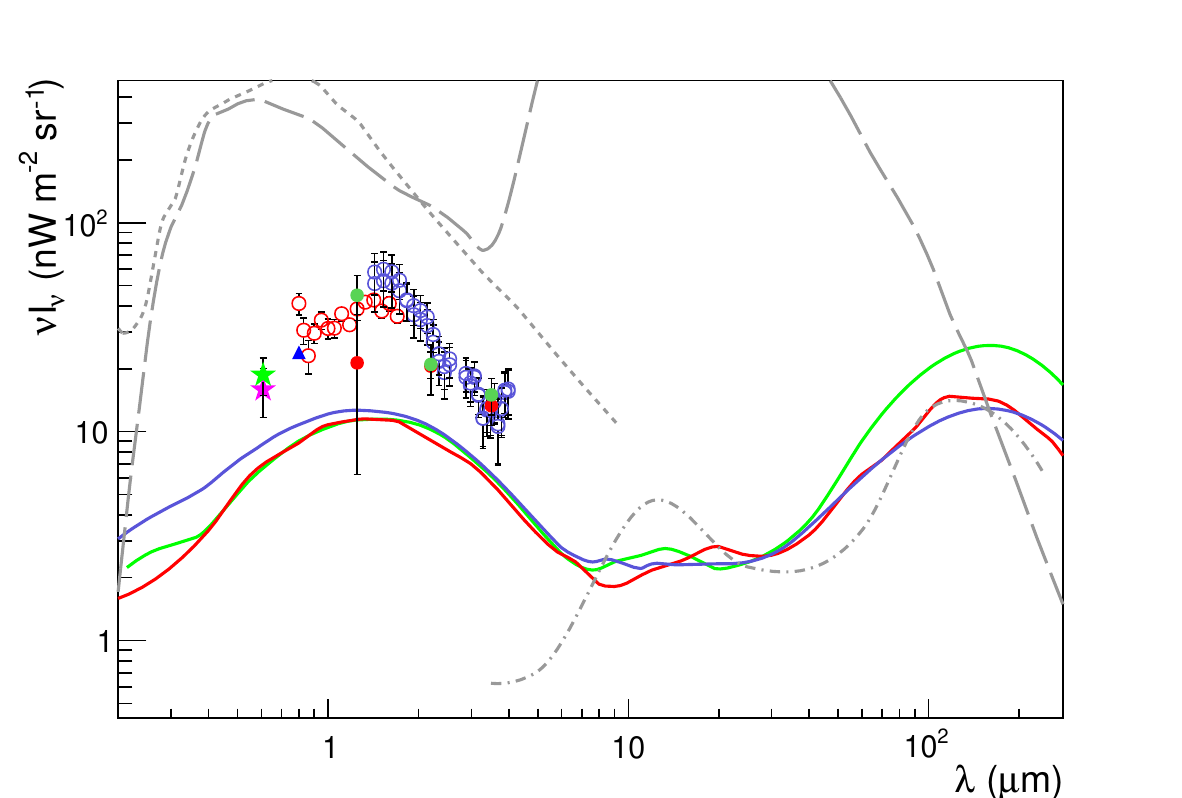}
    \caption{The predicted EBL intensity (z=0) versus wavelength by different models: \cite{fr2017} (red), \cite{Gilmore2012} (blue), \cite{dom2011} (green). Points correspond to recent EBL measurements in the optical-NIR band by various recent experiments. Open red circles: CIBER \citep{Matsuura2017}; open blue circles: IRTS renalaysis \citep{Matsumoto2015}; filled blue triangle: lower limit from the Hubble XDF \citep{Matsumoto2019}; filled red circle: DIRBE \citep*{levenson2007}; filled green circle: DIRBE \citep{Sano2020}; filled red and green stars: New Horizon \citep{Lauer2021}. The gray lines represent the major foregrounds \citep{leinert,Kashlinsky2005} in EBL measurements; dashed line: zodiacal light; dotted line: faint stars; dash-dotted line: cirrus.}
  \end{figure}

An excess population of EBL photon density (w.r.t. the estimated photon densities predicted by different EBL models and direct source counts) at $\sim\,1-4\,\mu$m was 
highlighted based on IRTS measurements \citep{Kashlinsky}. Even though \cite{Mattila2006} suggested this excess to be an artifact of improper 
zodiacal light subtraction, several subsequent measurements (by IRTS, HST, DIRBE, Spitzer, AKARI, New Horizons and CIBER) confirmed the excess -- 
at $\sim\,0.4-5\,\mu$m [\cite{Matsuura2017,Sano2020,Lauer2021}, and references therein]. This excess amounts to  several times the predicted 
EBL based on different EBL models. For example, DIRBE measured an intensity of $45\pm^{11}_{8}$ (at 1.25$\mu$m), $21\pm^{3}_{4}$ (at 2.2$\mu$m), 
$15\pm3$ (at 3.5$\mu$m) nWm$^{-2}$sr$^{-1}$, and CIBER indicates an intensity of $43\pm^{12}_{11}$\,nWm$^{-2}$sr$^{-1}$ (at 1.4$\mu$m) -- whereas 
model predictions are $\sim$1\,$\mu$m are $\sim$10\,nWm$^{-2}$sr$^{-1}$ [eg: \cite{Finke2010, Gilmore2012,fr2017, dom2011}]. A New Horizons LORRI 
measurement at a distance of $\sim$\,40\,AU from the Earth (where the zodiacal light is very faint) measured a diffuse flux component of intensity 
$8.8\pm 4.9$ to $11.9\pm 4.6$\,nWm$^{-2}$sr$^{-1}$ at 0.4-0.9\,$\mu$m,  which is a factor 2 above the predicted EBL.

The measured optical-NIR excess, a few times larger than estimations through galaxy counts { (see Fig.\,1)}, may be an artifact from  underestimating the zodiacal 
light \citep{Driver} -- but, on the contrary, it may be real and have an astrophysical origin. Population III (Pop-III) stars, which are primordial 
($z>6$), massive, metal-free stars, are a suitable candidate \citep{Matsumoto, Kashlinsky}. Considering the various  uncertainties in Pop-III parameters 
(i.e., redshift, star formation rate) the estimation of their integrated emission is challenging. For example, for the optical peak \cite{Salvaterra} 
estimated $\sim$ 60\,nWm$^{-2}$sr$^{-1}$  whereas \cite{Sun}, obtained { a model estimate of} $<$0.2\,nWm$^{-2}$sr$^{-1}$. In the latter 
case, any signature of Pop-III stars in the EBL will be hardly noticed.

{ Decay of axion or axion-like particles (ALP)}  is another candidate to explain the HST and CIBER measurements between 0.6 and 1.6\,$\mu$m, provided the axion mass is within 
$\sim$1-10\,eV \citep{Gong}. Direct collapse of a population of Compton-thick high-$z$ black holes may also result in an excess EBL  intensity 
of $\sim 1\,{\rm nW}m^{-2}sr^{-1}$ at $\sim 2\mu$m  \citep{Yueb}.
By using the LIGO black hole parameters \cite{Kashlinsky2} argued that a sufficiently large number of primordial black holes can result in an 
additional EBL source population at 2-5\,$\mu$m. Recently, \cite{Lauer2021} justified the  New-Horizons--measured unresolved EBL component 
invoking undetected $z>6$ galaxies (assuming a steeper galaxy-count slope at V$>$24, or a missing population of galaxies with V$<$30).

After ruling out the arguments  on the excess photons as originating from zodiacal light, diffuse Galactic light, known galaxies clustering, 
primordial galaxies,  or black holes during cosmic reionization, \cite{Zemcov} associated the extra emission with intra-halo light (IHL), 
i.e. the emission from tidally stripped stars from their host galaxies at $z<3$. { On the contrary} the spatial fluctuations of HST sky surveys at $0.6 < \lambda/\mu{\rm m} < 1.6$, \cite{Mitchell-Wynne} dispense with the invoked extra populations of IHL-producing galaxies and $z>6$ galaxies. 

Hence, combining observations (from different instruments and background models) and phenomenological estimations, there clearly is an ambiguity 
in the optical-NIR band of the EBL, that leaves a possibility to accommodate an additional photon component of astrophysical origin. In what follows, 
we study this possibility { using constraints derived by correcting observed VHE $\gamma$-ray spectra for the absorption due to an EBL model that 
consists of a baseline component (arising from galaxy counts) and an additional hypothetical component (of unspecified astrophysical origin). }

\begin{table}
	\centering
	\label{tab:example_table}
	\begin{tabular}{lccr} 
		\hline
		AGN & Redshift & Reference\\
		\hline
		Mrk\,421 (12 spectra) & 0.031  & 1 - 7\\
		Mrk\,501 (17 spectra) & 0.034  & 8 - 16\\
		1ES\,2344+514 (2 spectra) & 0.044  & 17, 18\\
                Mrk\,180 & 0.045 & 19\\
                1ES\,1959+650 (5 spectra) & 0.048 & 19 - 21\\
                AP\,Librae & 0.049 & 22\\
                1ES\,2037+521 & 0.053 & 23\\
                PKS\,0625-35 & 0.055 & 24 \\
                1ES\,1727+501 & 0.055 & 25 \\
                PGC\,2402248 & 0.065 & 26 \\
                PKS\,0548-322 & 0.069 & 27\\
                BL\,Lacertae (5 spectra) & 0.069 & 28 - 31\\
                PKS\,2005-489 (2 spectra) & 0.071 & 32, 33\\
                RGB\,J\,0152+017 & 0.080 & 34\\
                W\,Comae & 0.102 & 35\\
                1ES\,1312-423 & 0.105 & 36\\
                RGB\,J\,0521.8+211 & 0.108 & 37\\
                PKS\,2155-301 (13 spectra) & 0.116 & 38 - 42\\
                RGB\,J\,0710+591 & 0.125 & 43\\
                H\,1426+428 & 0.129 & 44\\
                1ES\,1215+202 (2 spectra) & 0.131 & 45, 46\\
                1ES\,0806+524 (3 spectra) & 0.138 & 47, 48\\
                1ES\,0229+200 & 0.140 & 49\\
                1RXS\,J\,101015.9-311909 & 0.143 & 50\\
                H\,2356-309 (4 spectra) & 0.165 & 51, 52\\
                RX\,J\,0648.7+1516 & 0.179 & 53\\
                1ES\,1218+304 (3 spectra) & 0.182 & 54 - 56\\
                1ES\,1101-232 (2 spectra) & 0.186 & 57, 58\\
                1ES\,0347-121 & 0.188 & 59\\
                RBS\,0413 & 0.190 & 60\\
                1ES\,1011+496 (3 spectra) & 0.212 & 61 - 63\\
                1ES\,0414+009 (2 spectra) & 0.287 & 64, 65\\
                PKS\,1510-089 (5 spectra) & 0.361 & 66 - 68\\
                PKS\,1222+216 (2 spectra) & 0.432 & 69, 70\\
                3C\,279 & 0.536 & 71\\
                B2\,1420+32 &0.682 & 72\\
                PKS\,1441+25 (3 spectra) & 0.939 & 73, 74\\
		\hline
	\end{tabular}
        
	\caption{
	  The data set used for this study. The columns are, the source name, redshift and reference numbers. The corresponding references are as follows:
          1\citep{421a} 2\citep{421c} 3\citep{421d}  4\citep{421l} 6\citep{421m} 7\citep{421n} 8\citep{501a}  9\citep{501d} 10\citep{501e} 11\citep{501g} 12\citep{501i} 13\citep{501j} 14\citep{501k} 15\citep{501n} 16\citep{2344a} 17\citep{2344b} 18\citep{180a} 19\citep{1959a} 20\citep{1959b} 21\citep{1959c} 22\citep{aplib} 23\citep{2037} 24\citep{0625} 25\citep{1727} 26\citep{pgc} 27\citep{0548} 28\citep{bla} 29\citep{blb} 30\citep{blc} 31\citep{bld}
          32\citep{2005a} 33\citep{2005b} 34\citep{0152} 35\citep{wcom} 36\citep{1312} 37\citep{0521} 38\citep{2155c} 39\citep{2155e} 40\citep{2155l} 41\citep{2155o} 42\citep{2155p} 43\citep{0710} 44\citep{1426} 45\citep{1215a} 46\citep{1215b} 47\citep{0806a} 48\citep{0806b} 49\citep{0229} 50\citep{101015} 51\citep{2356a} 52\citep{2356b} 53\citep{0648} 54\citep{1218a} 55\citep{1218b} 56\citep{1218c} 57\citep{1101a} 58\citep{1101b} 59\citep{0347} 60\citep{0413} 61\citep{1011a} 62\citep{1011b} 63\citep{1011c} 64\citep{0414a} 65\citep{0414b} 66\citep{1510a} 67\citep{1510b} 68\citep{1510c} 69\citep{1222a} 70\citep{1222b} 71\citep{3c279} 72\citep{1420} 73\citep{1441} 74\citep{1441b}
                  }
\end{table}

 \section{Data sets}

We have collected 105 VHE spectra from  the major Imaging Atmospheric Cherenkov Telescopes (IACTs) in current operation, H.E.S.S. \citep{hesst}, 
MAGIC \citep{magict}, and VERITAS \citep{veritast}. These spectra correspond to 37 blazars and span the redshift range $0.03 < z < 0.94$. We limit 
our sample such that each spectrum should have at least 4 spectral points (in order to apply the method described below, see Section\,4). Also, 
we do not include sources with unknown, { indirect} or disputed redshifts (e.g., PG\,1553+113, 3C 66\,A, PKS 1424+240, S5 0716+714). Considering their close 
proximity (hence, the negligible effect  of the EBL on their observed spectra), the radio galaxies Centaurus\,A and M\,87 are also not 
considered. The final sample of VHE spectra used in this study is given in Table\,1. In { Fig.2} we plot the datapoint energies against source 
redshift, and compare them with the optical depth of  the \cite{fr2017} EBL model. 

 \begin{figure}
   \hspace{-0.5cm}
   \includegraphics[width=1.2\columnwidth]{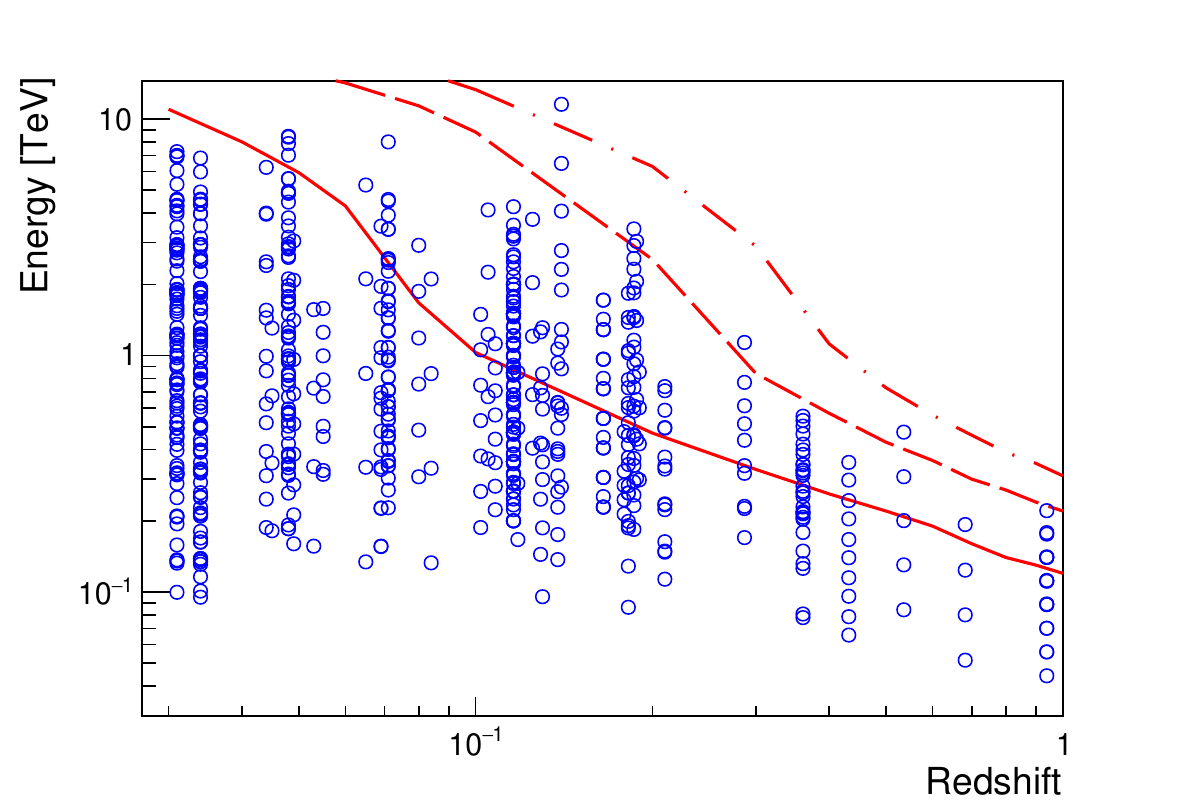}
   \vspace{-0.5cm}
   \caption{The  energy vs. redshift distribution of the VHE spectral points considered in this study.
     The solid, dashed, and dot-dashed lines respectively correspond to the optical depths of $\tau=$1, 3, and 5 \citep{fr2017}.}
    \label{fig:example_figure}
\end{figure}


\section{Method}

\begin{figure*}
  \setcounter{figure}{2}
        \centering
        
      \begin{multicols}{2}
        \subcaptionbox{\label{figu:3}}{\includegraphics[width=1.15\linewidth]{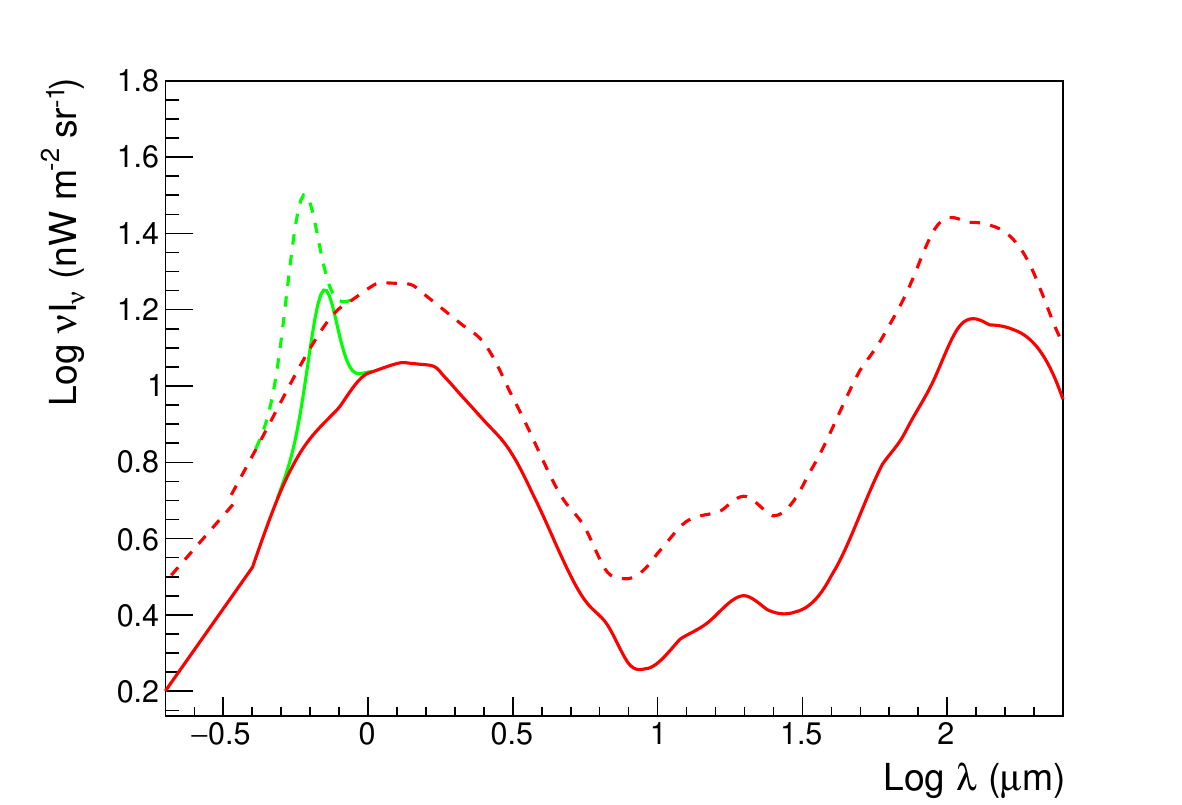}}\par 
        \subcaptionbox{\label{figh:4}}{\includegraphics[width=1.15\linewidth]{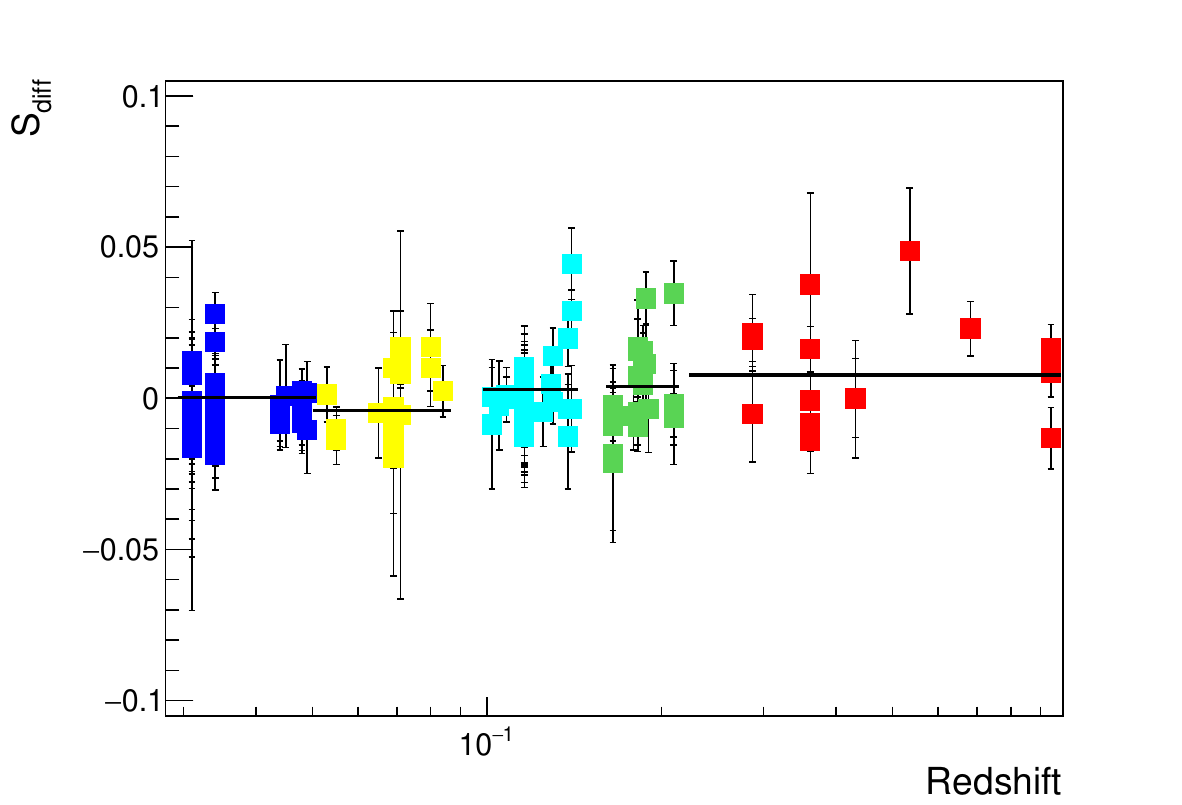}}\par 
        \end{multicols}
      \begin{multicols}{2}
            \subcaptionbox{\label{figl:5}}{\includegraphics[width=1.15\linewidth]{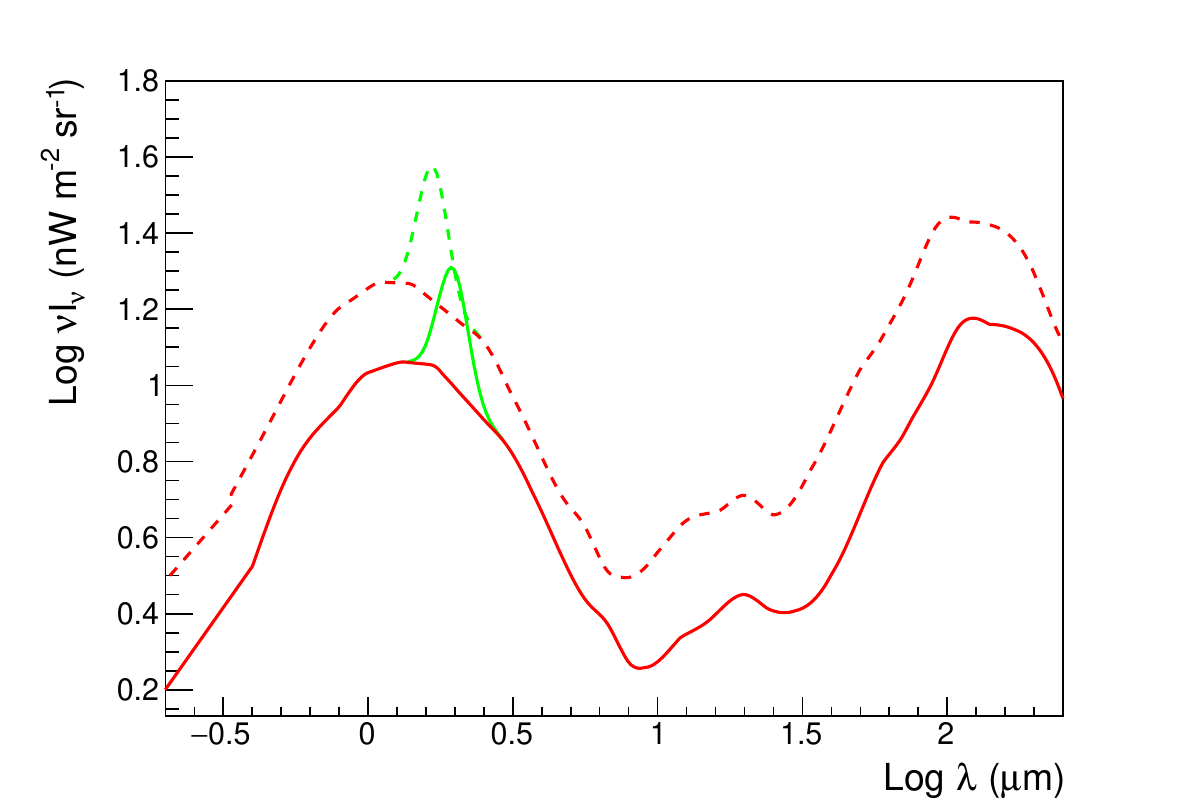}}\par 
            \subcaptionbox{\label{figm:6}}{\includegraphics[width=1.15\linewidth]{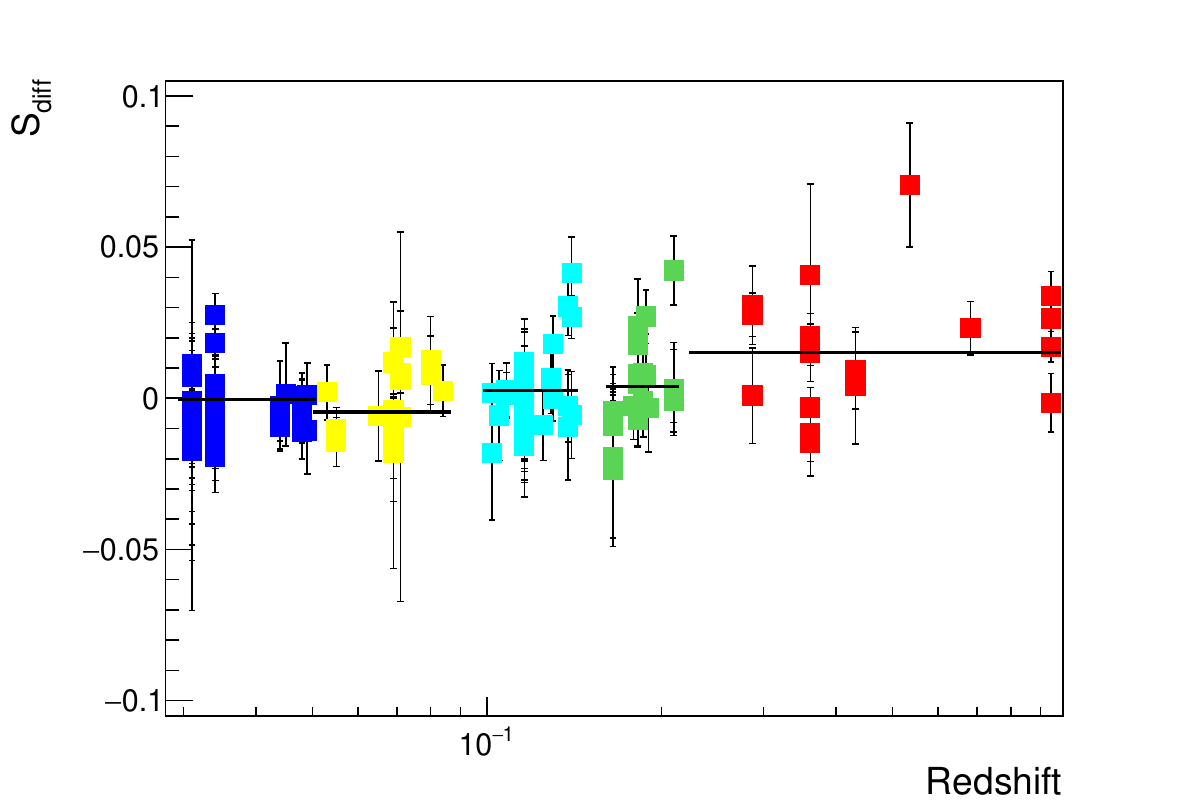}}\par 
        \end{multicols}
      
    \caption{ The baseline \cite{fr2017} EBL model (red solid line), and the modified (baseline plus additional) EBL model (green solid line) at $z=0$ 
             in the optical-NIR band. The dashed lines represent the corresponding EBL at $z=0.6$. 
             {\it Left:} the additional component, of amplitude 10\,$nWm^{-2}sr^{-1}$, is located at 0.71\,$\mu$m ({\it top}) and 2.0$\mu m$ {\it bottom}. 
             {\it Right:} the corresponding distribution of the  normalized difference S$_{\rm diff}$ (see Eq.\,2) for the various redshift bands are shown ($0.00< z \leq 0.05$: blue; $0.05 < z \leq 0.10$: yellow; $0.10< z \leq 0.15$: cyan; $0.15< z \leq 0.20$:green; $z>0.25$: red.}
    
    \end{figure*}

We choose the recent \cite{fr2017} EBL model as our baseline model. This  provides the  {\it guaranteed} EBL density from galaxy number 
counts and luminosity functions. We further add different sets of Gaussian-shaped population components to the baseline model { in  
$\rm{log\left(\lambda\right)}$-$\rm{log\left(\nu I_{\nu}\left(z\right)\right)}$  space (where $\lambda$ is the { observed}
EBL photon wavelength in $\mu$m, and  $\nu I_{\nu}\left(z\right)$  is the EBL intensity  { at redshift $z$} in nW\,m$^{-2}$sr$^{-1}$)} at the 
optical-NIR peak, i.e. 
\begin{equation}
  {  \rm{F}_{\rm{add}}= B\left(z\right)\, \rm{exp}\left[-\frac{(x\left(z\right)-\mu\left(z\right))^2}{2\sigma^2}\right]\,,}
	\label{eq:gauss}
\end{equation}
{ where $B\left(z\right)$ and $\mu\left(z\right)$ are normalization and mean wavelength ($\lambda_{\rm m}$)} in logarithmic scale of the additional component distribution, 
$z$ is the redshift, $\sigma$ 
is the standard deviation, and $x = {\rm log}(\lambda)$. We assume { that} the  number density of this extra population increases with redshift as 
$\left(1+z\right)^3$. { This factor takes care of the expansion of the Universe, under the assumption that the photon field originates from higher 
redshift than the VHE sources we considered. Since the astrophysical photon fields discussed in Section 2 are of high-$z$ origin, this is a 
reasonable assumption.} To obtain optical depths at different spectral energies and redshifts, the resulting photon number densities 
(combination of the baseline EBL model and the Gaussian function population) are then integrated using Eq.(13) of \cite{fr2008}.

%

\begin{figure*}
  \setcounter{figure}{3}
        \centering
        \begin{multicols}{2}
          \subcaptionbox{\label{fig:1}}{\includegraphics[width=1.05\linewidth]{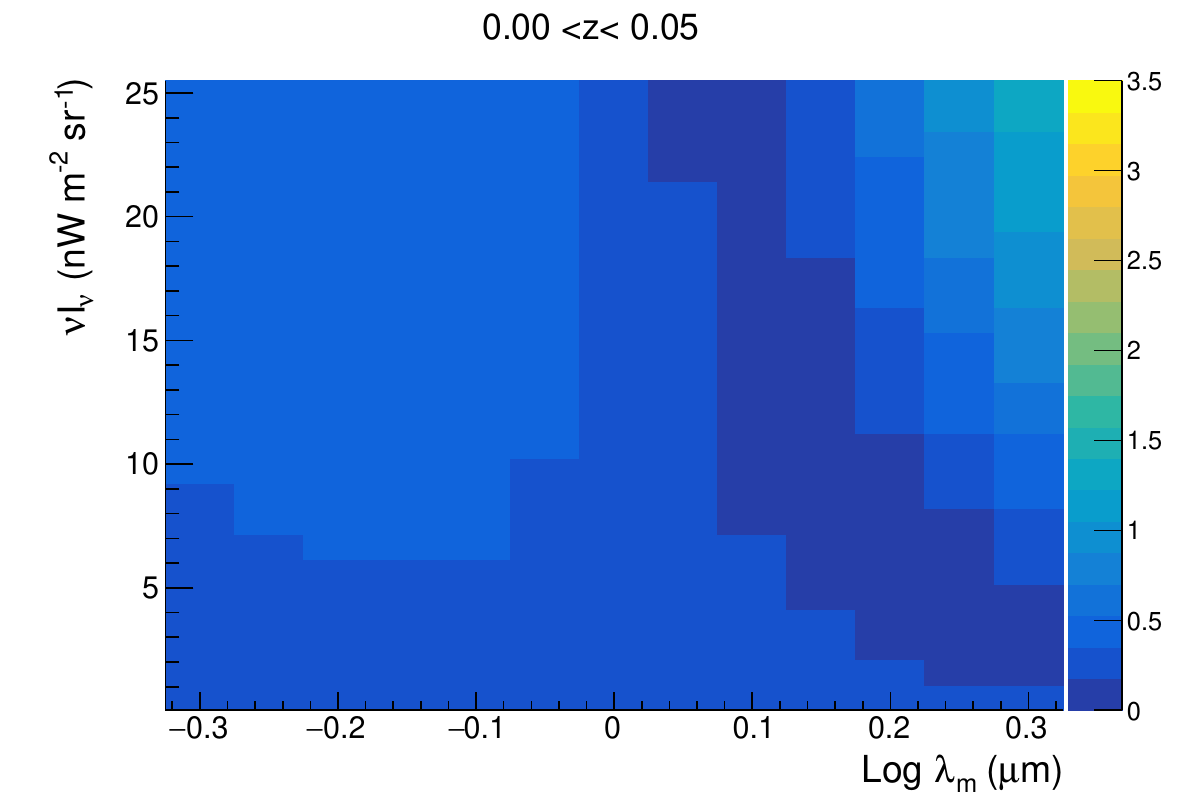}}\par
          
            \subcaptionbox{\label{fig:2}}{\includegraphics[width=1.05\linewidth]{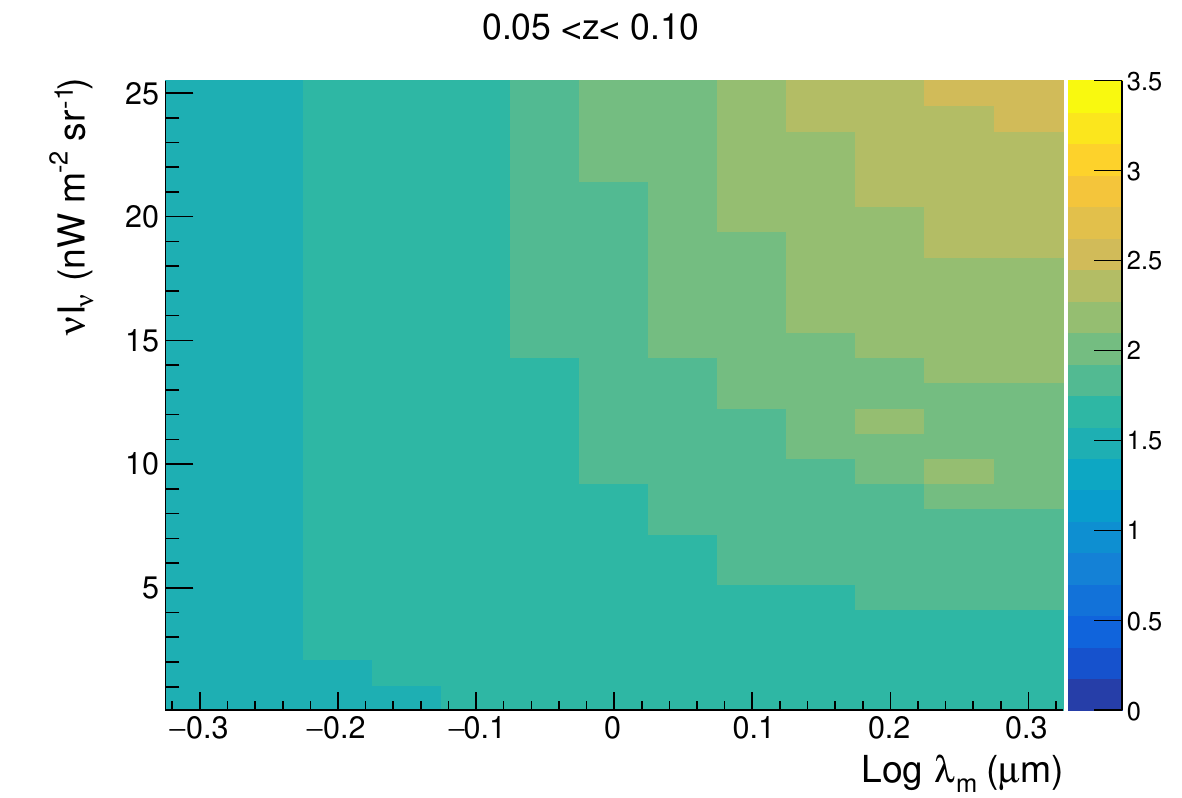}}\par 
        \end{multicols}
        \begin{multicols}{2}
            \subcaptionbox{\label{fig:3}}{\includegraphics[width=1.05\linewidth]{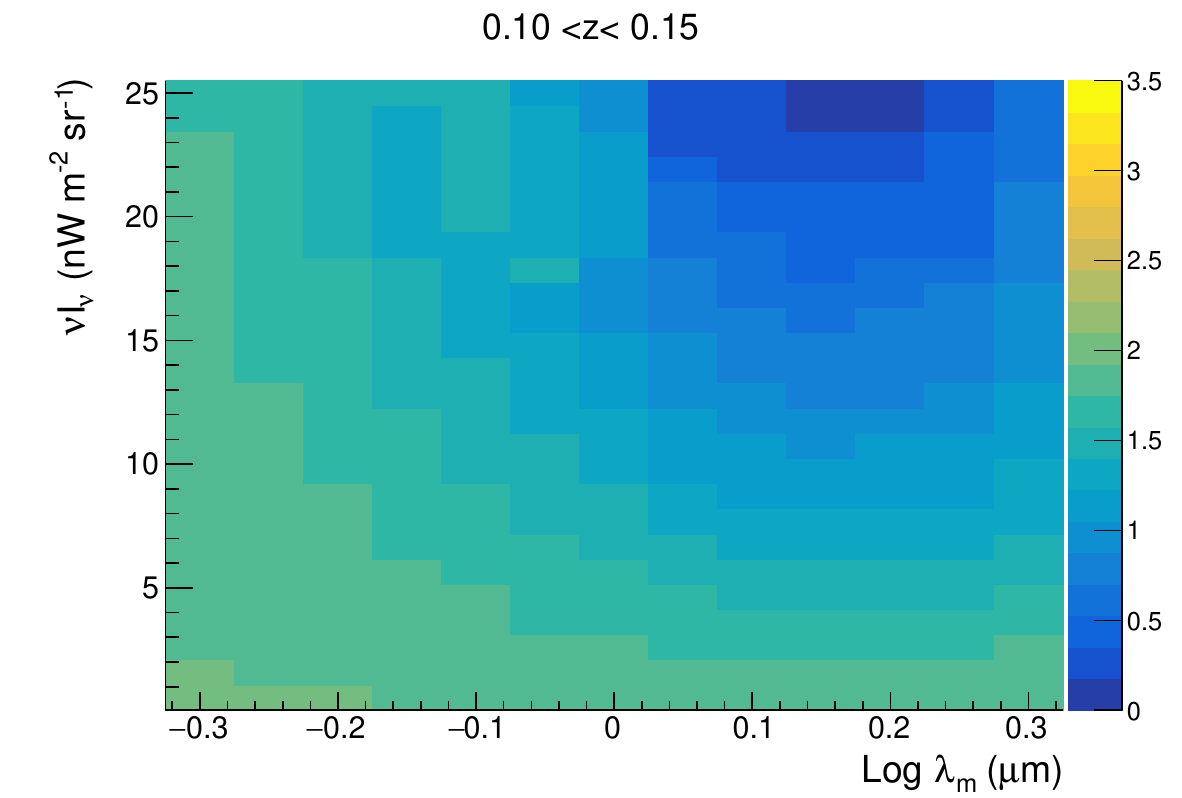}}\par
            \subcaptionbox{\label{fig:4}}{\includegraphics[width=1.05\linewidth]{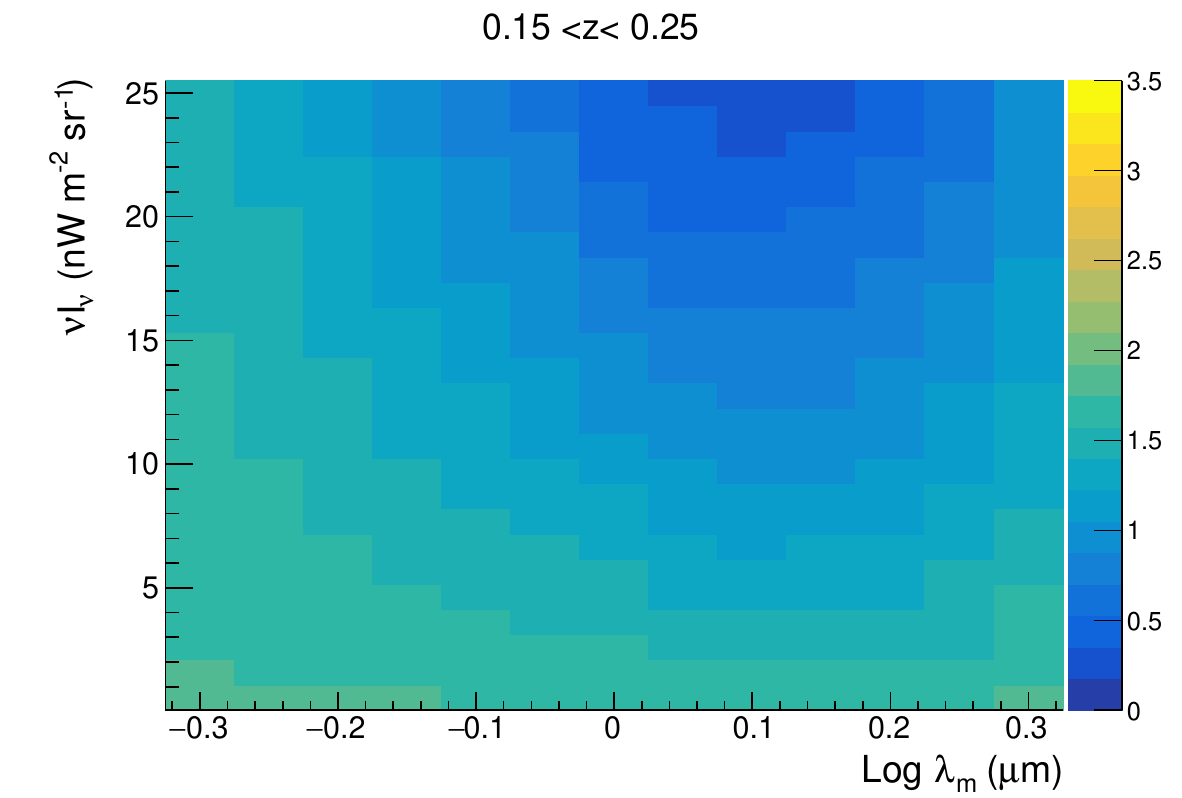}}\par
        \end{multicols}
        \begin{multicols}{2}
            \subcaptionbox{\label{fig:3}}{\includegraphics[width=1.05\linewidth]{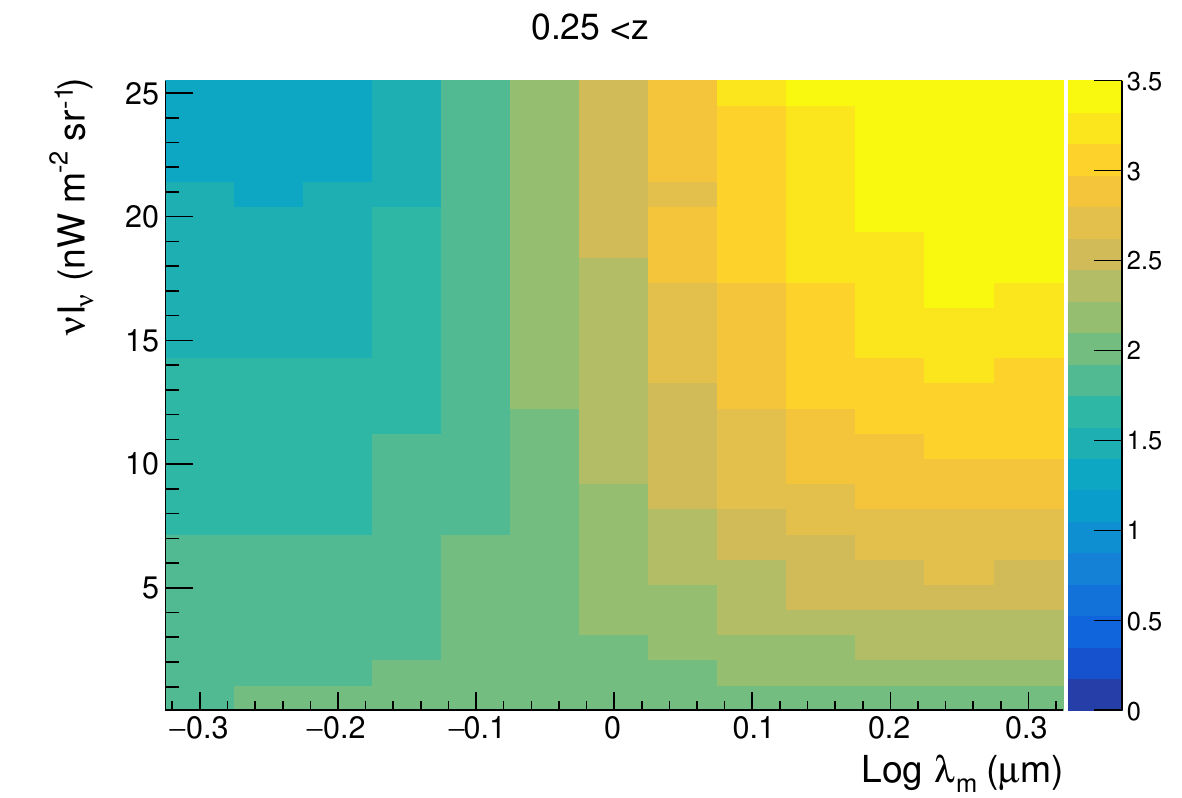}}\par
            \subcaptionbox{\label{fig:4}}{\includegraphics[width=1.05\linewidth]{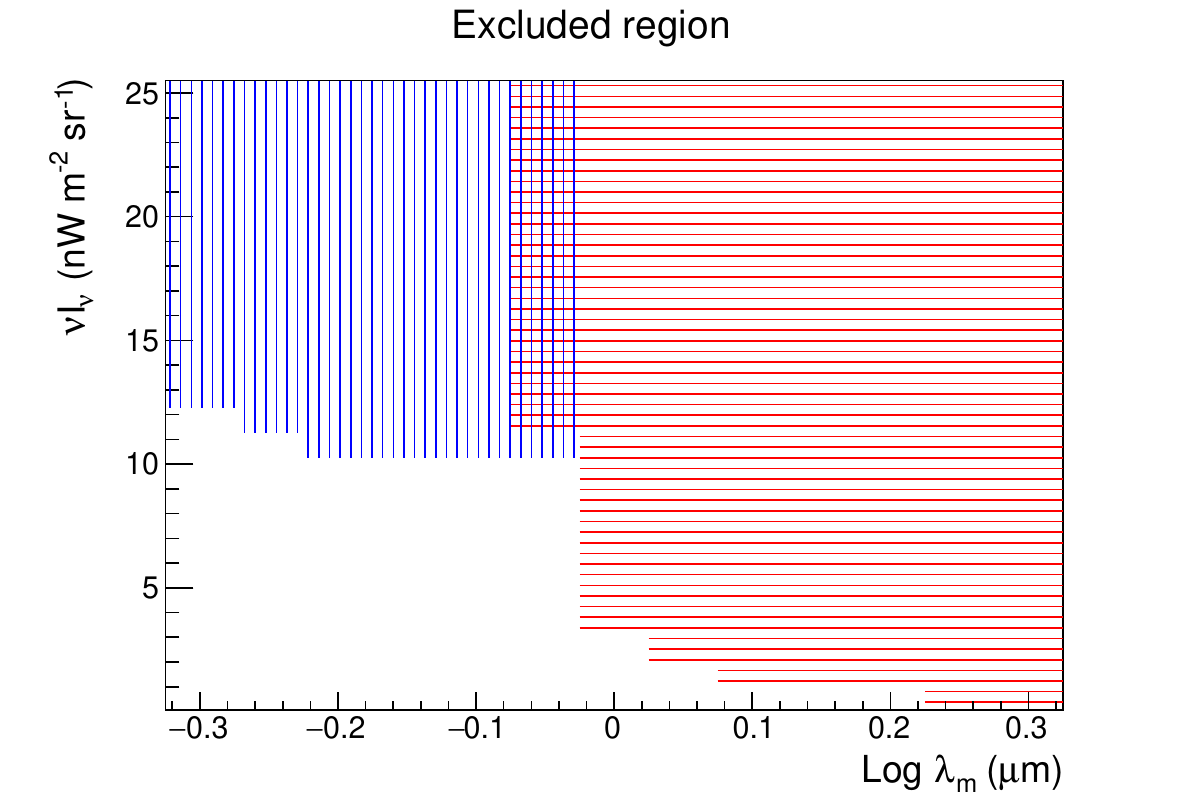}}\par
        \end{multicols}
    \caption{ 2-D histograms (z=0) showing the t-value distribution when the additional 
component corresponds to different values of $\mu$ and B.  The colours correspond to the t-values of each bins (see text also). The $\sigma$ used in these plots is 0.05. 
The bottom-right figure is the combined image of all histograms. The red horizontal stripes region 
corresponds to p-value more than $t_{95}$, hence excluded based on the t-test. The blue vertical 
stripes correspond to the excluded region based on the spectral index  $\alpha <$\,1.5 of 
any spectrum. The white non-striped region is the resulting acceptable zone for $\sigma=$0.05.}
    \end{figure*}

In order to segregate the acceptable additional population that can be accommodated on top of the baseline model, we scan the 
$B$, $\mu$ parameter space for a fixed value of $\sigma$. This operation is repeated for several values of $\sigma$. (\cite*{Neronov} 
adopted a similar approach employing $\gamma$-ray spectra of 7 blazars.) 

As to the acceptance criteria for the excess populations, we adopt two approaches  common in EBL studies when using VHE spectra. 
The first approach  relies on the upward curvature of the de-absorbed VHE spectra of blazars \citep{mazin1553,meyer}. Considering 
synchrotron radiation in the optical/X-ray band and Synchrotron-Self-Compton and External Compton emission in the $\gamma$-ray band, 
concave VHE spectra are very unlikely. In order to quantify the  spectral shape, we apply the new sets of EBL densities (i.e. the 
baseline model plus different extra populations) and fit the lower-energy spectrum (see next sentence) with a power-law shape 
($\phi(E) \propto (E/E_0)^{-\alpha}$) or a Log Parabola shape ($\phi(E) \propto (E/E_0)^{-\alpha-\beta\, \rm{ln}(E/E_0)}$), in which 
$E_0=\sqrt(E_{min}E_{max})$ where $E_{min}$ and $E_{max}$ are the minimum and maximum photon energy in the VHE spectrum, $\alpha$ is 
the photon index, and $\beta$ is the curvature parameter.
 We have then  selected the fit which favours the lower reduced $\chi^{2}$ values. We have also checked the dependence of the parameter E$_0$ by arbitarily choosing different values, and found that this does not exert an appreciable difference to our results. More complex intrinsic spectral shapes (eg: power-law with an exponential cut off) were also considered. However, the F-test did not yield a statistically significant (95\%) improvement of the fit results in such cases. Hence, complex functions (with more parameters) were not used in this study.

We extrapolate the fitted spectrum to the highest-energy point(s) -- which 
 was (were) not used for the fit. Such points are chosen as the last single point if the spectrum has $\leq$5 spectral points, 
or the last two points if the spectrum has $>$5 points. 
We then estimate the difference between the flux at { the} highest-energy spectral point and the corresponding extrapolated flux. To avoid 
any bias arising from spectra of different flux levels, we normalize the differences obtained  according to 
\begin{equation}
\rm{S_{diff}}=\frac{F_{int}-F_{ext}}{F_{ext}+F_{int}}
   \label{eq:gaus}
\end{equation}
where $\rm{F_{ext}}$ is the extrapolated flux and $\rm{F_{int}}$ is the intrinsic flux at the highest energies. The resulting distribution 
after sampling $\rm{S_{diff}}$ in bins of different redshifts should ideally spread around 0 in the absence of any bias. The $\rm{S_{diff}}$ 
values from all the datapoints in different redshift bins are compared using the 'Student's T statistical test'(t-test). The null hypothesis 
for the t-test is that the differences are statistically not significant. If the null hypothesis is rejected, the corresponding EBL population 
(i.e., the baseline model plus the extra population) is ruled out.

{ The second approach relies on examining the derived intrinsic blazar VHE spectra in the light of standard particle acceleration  constraints in both the leptonic and hadronic scenarios, i.e. the intrinsic spectral index is expected to be $\rm{\alpha} > 1.5$ within error bars \citep{2356,mazinebl}. Hence, we reject any extra component that gives rise to $\rm{\alpha} < 1.5$, for any of the spectra.}

\section{Results}


The t-test values are calculated for the $\rm{S_{diff}}$ distribution in the redshift intervals 0.0-0.05, 0.05-0.10, 0.10-0.15, 0.15-0.25, 
and $>$0.25 for several additional Gaussian-shaped EBL components. 
We first estimate the t-values of the distribution without adding any extra components. The t-values are 0.6, 1.6, 1.9, 1.8, and 1.6 with 
number of points 65, 20, 41, 25 and 17 in the reshift bands  0.0-0.05, 0.05-0.10, 0.10-0.15, 0.15-0.25, and $>$0.25. These values are below 
the 95\% confidence level (i.e., t-values 2.0, 2.09, 2.02, 2.06, and 2.11 respectively). This indicates 
no significant evidence against the null hypothesis, i.e., the differences $\rm{S}_{\rm{diff}}$ are not statistically significant (within 
95\% confidence level). We then calculate the t-test values after adding different Gaussian extra-populations to the baseline model.


{ For this, $\mu$ varies (by steps of 0.05) within -0.30$-$0.30 (corresponding to wavelengths $\lambda \sim$ 0.5$-$2.0 $\mu$m), and B varies by steps of 1 ${\rm nW}m^{-2}sr^{-1}$} { (after setting $\sigma$ to  a constant value}
\footnote
{ 
Since the Gaussian spectrum is generated at high redshifts and photon energies 
and frequencies both scale as $(1+z)^{-1}$, $\sigma$ is independent on redshift.
})
{ in order to obtain a { total (model)} EBL intensity at the optical-NIR peak of $\lesssim$35 ${\rm nW}m^{-2}sr^{-1}$ 
}

An example of $\rm{S_{diff}}$ distribution is given in { Fig.\,3\,(b, d), with the corresponding EBL (Fig\,3\,a, c).} The t-values obtained 
for each such combinations (with floating $\mu$ and B, and fixed $\sigma= 0.05$) at different redshift ranges are shown as 2-D histograms in 
{ Fig\,4\,(a, b, c, d, e)}. The obtained t-values in the lower-redshift samples do not significantly influence the rejection of extra-population 
EBL models. However, the t-value distributions for the high-redshift (z$>$0.25) samples clearly tell the acceptable from the forbidden region 
(i.e., some specific extra-populations EBL models are not supported) in the $\mu$-B plot. This is because of the strong cosmological increase, 
$\propto (1+z)^3$, of the additional population. { Fig\,4\,(f)} shows a combined image of the forbidden zone (with $\sigma=$0.05) in our analysis: 
the horizontal red strips region shows the region excluded at 95\% confidence level based on Student's t-test. We then check all the 
EBL-deabsorbed spectra for each set of extra-populations: models with $\mu$ and B values leading to a spectral index $<$1.5 are ruled out: 
their region is denoted with vertical blue lines in { Fig\,4\,(f)}. Hence, the acceptable combinations of $\mu$ and B (for $\sigma=$0.05) are 
found in the non-shaded region of the figure.

\begin{figure}
  \setcounter{figure}{4}
  \vspace{1.cm}
  \hspace{-0.5cm}
   \includegraphics[width=1.2\columnwidth]{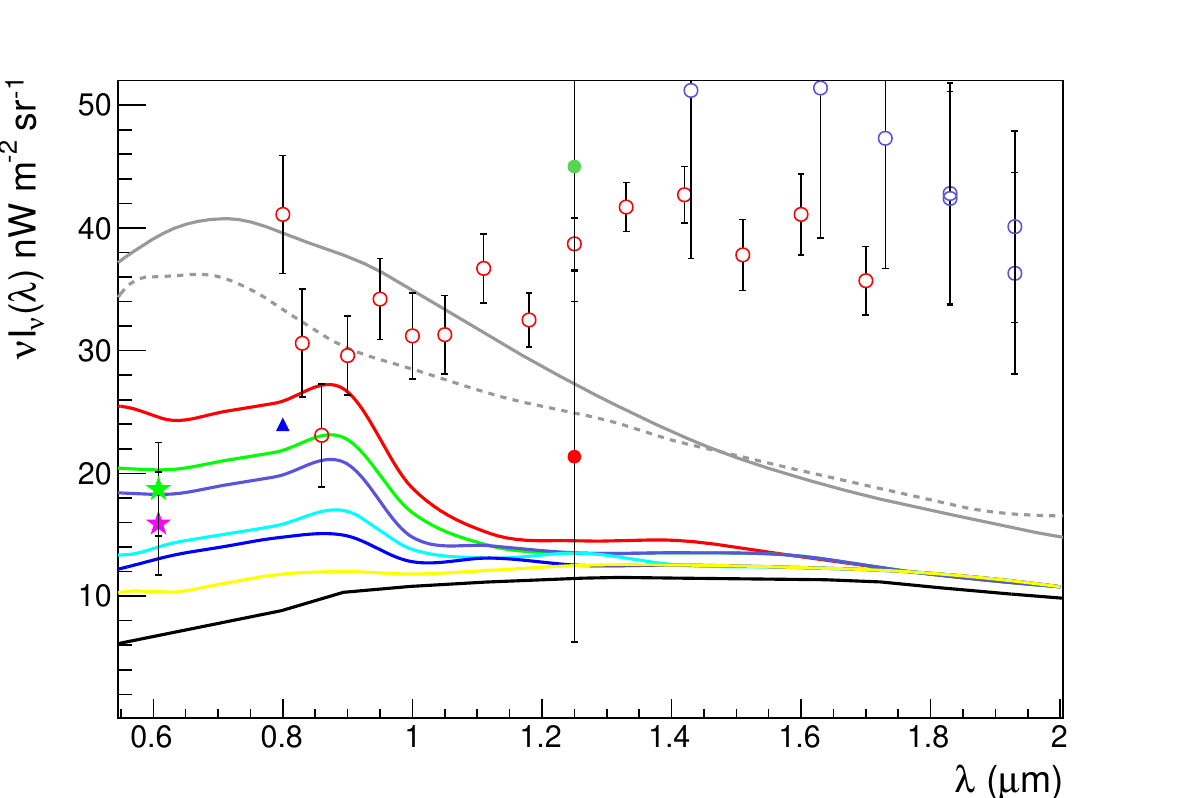}
   \caption{The obtained  95\% confidence level upper limits  on the amplitude of the  Gaussian component { (superposed on  the baseline model)} against optical-NIR wavelengths is shown 
as continuous lines. The red, green, violet, cyan, blue, and yellow lines correspond to  the cases $\sigma$=0.03, 0.04, 0.05, 0.07, 0.09, and 0.12, 
respectively. The black line denotes the baseline EBL model. Points representing EBL measurements are as in Fig. 1. We also show the upper limits obtained by \cite{mazinebl} and \cite{meyerebl} in solid and dashed gray lines respectively.}
\label{fig:example_figure}
\end{figure}

Next we repeat the exercise considering  Gaussian components with different $\sigma$'s starting from 0.03 (this can be considered very 
narrow, and corresponds to $\sim$4\% of the $\sigma$ of the baseline-EBL's optical peak if  the latter is roughly approximated with a 
Gaussian function) to 0.12, by intervals of 0.01. Clearly, the larger is $\sigma$ the lesser  is the possibility of adding an 
extra-population with a large amplitude to the existing EBL. For example, there is no  room for a significant ($\gtrsim1$ nW$m^{-2}sr^{-1}$) 
Gaussian component for $\lambda > 0.89\,\mu$m if $\sigma=$0.12. However,  such a component with $\sigma=$0.03 is permitted for $\lambda \leq 1.4\,\mu$m. Similarly, only an extra-population of amplitude 1.1\,nW$m^{-2}sr^{-1}$ can be accommodated at 0.79\,$\mu$m for $\sigma=0.12$, 
whereas a population contributing $\leq$18.7\,nW$m^{-2}sr^{-1}$ at the same $\lambda$ is admissible for $\sigma=0.03$.

{
 In Fig.\,5 we show the obtained upper limits for the amplitude of the Gaussian component 
(Y-axis) if they were placed at different wavelengths (X-axis).
As shown in the figure, assuming a narrow ($0.03<\sigma<0.12$) excess Gaussian components we rule out, 
with a confidence level of $95\%$, any significant (i.e., peak amplitude $>1\,{\rm nW}\,m^{-2}sr^{-1}$) additional population at $\lambda >$ 1.5 
$\mu$m.
}

\begin{figure*}
  \setcounter{figure}{5}
        \centering
        \begin{multicols}{2}
          \subcaptionbox{\label{fig:1}}{\includegraphics[width=1.1\linewidth]{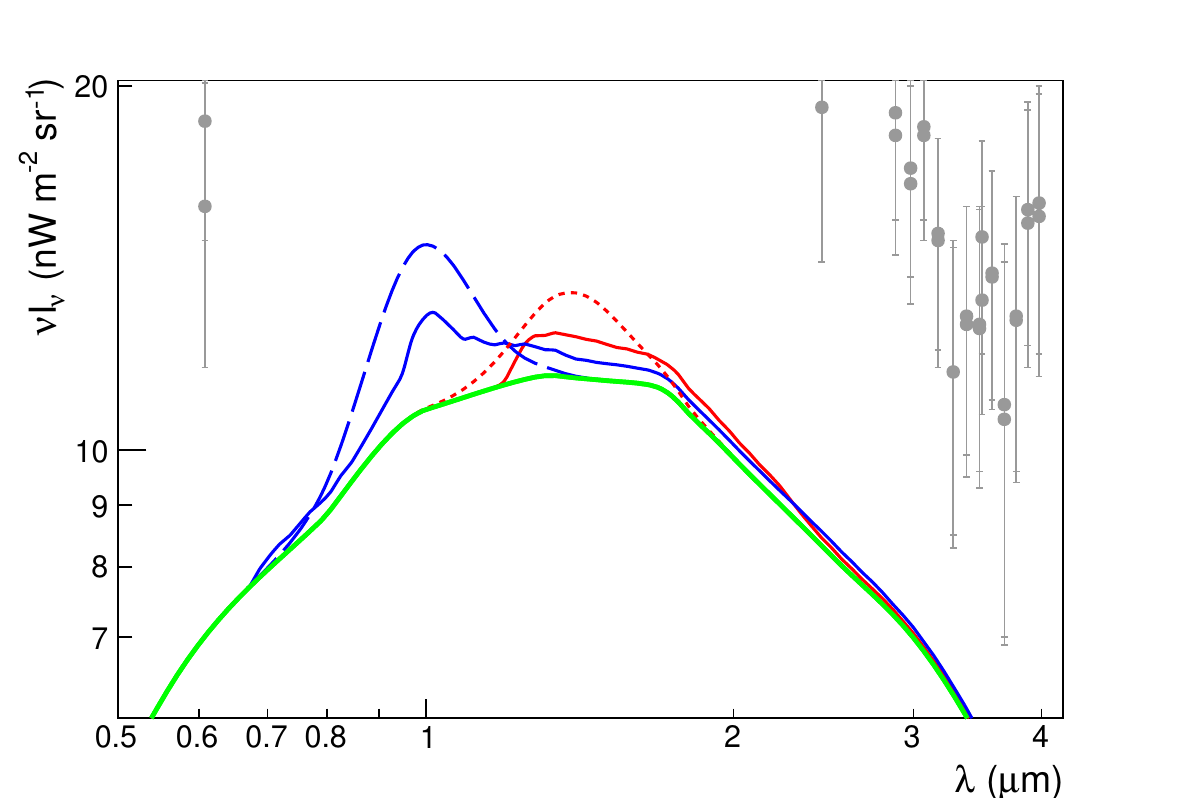}}\par
          
            \subcaptionbox{\label{fig:2}}{\includegraphics[width=1.1\linewidth]{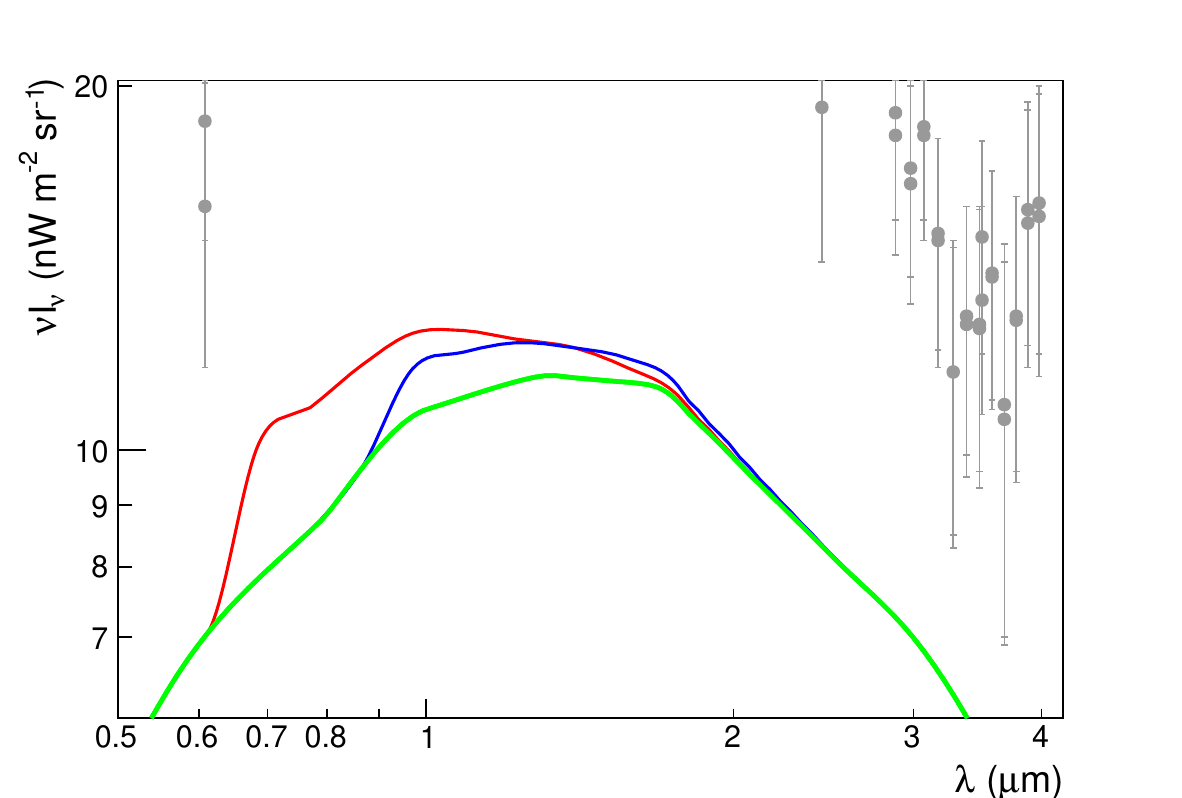}}\par 
        \end{multicols}
        
    \caption{ Examples of acceptable Pop-III models (on the left), and ALP annihilation models (on the right) together with the baseline EBL model (green solid line). The direct measurements below 20\,nW\,m$^{-2}$sr$^{-1}$ are shown in gray (see Fig.\,1 for details). (a) The red line represents the modified \citep{Salvaterra} model after scaling the original star formation efficiency down by a factor of 60, while the blue line shows the \cite{Fernandez2010} model after scaling the star formation efficiency down by a factor of 2. The dashed red and blue lines are the upper limits of Gaussian contribution (of $\sigma$=0.05) for comparison. Considering the wider distribution of the Pop-III model populations, the spectrum would be lower than Gaussian ULs. (b) Maximum allowed ALP annihilation component in the case of ALP mass $m_{a}$=2.8\,eV (blue), and $m_{a}$=4.0\,eV (red).}
    \end{figure*}

\section{Discussion}


Our results {are in conflict with} recent estimations (at $\lambda \gtrsim$ 1.2\,$\mu$m) by IRTS \citep{Matsumoto2015}, DIBRE \citep*{Sano2016,Sano2020}, 
AKARI \citep{Tsamura2013}, and CIBER \citep{Matsuura2017}. Our estimations also differ from the HST measurements \citep{Kawara2017} at { shorter} 
wavelengths ($<1\,\mu$m), that anticipate an excess at a level of a few times the obtained upper limits. Recent measurements carried out by LORRI onboard the New Horizons satellite at a distance $>$40 AU from the Sun yielded an EBL intensity of 
15.9$\pm$4.2 to 18.7$\pm$3.8 nWm$^{-2}$sr$^{-1}$ at 0.608 $\mu$m \citep{Lauer2021}. After subtracting contributions from known sources, the 
remaining excess is 8.8$\pm$4.9 to 11.9$\pm$4.6 nWm$^{-2}$sr$^{-1}$, in agreement with our estimates assuming excess components 
{ ($\sigma \lesssim 0.1$)}. The earlier LORRI estimates, that used data from limited observation time at $\sim$1.9\,AU distance from the Sun \citep{Zemcov1}, 
also provided a 2$\sigma$ upper limit on the EBL, 19.3 nWm$^{-2}$sr$^{-1}$ at 0.65 $\mu$m, which is well within our estimates. In { Fig.\,5} we also
compare our estimates with { recent} direct measurements.

It is interesting to note that the results obtained from large samples of VHE spectra \citep{mazinebl,meyerebl} are also in good agreement with 
our estimates, as they  show a clear distinct peak at $\sim$ 0.7\,$\mu$m.  \cite{mazinebl} predicts an upper limit of 40\,nWm$^{-2}$sr$^{-1}$, 
while \cite{meyerebl} anticipates an upper limit of 24\,nWm$^{-2}$sr$^{-1}$. On the other hand, an additional peak was obtained by \cite{Neronov} { by} 
employing the $\gamma$-ray spectra of 7 relatively nearby AGNs. The peak located at $\lambda = 1.7\mu$m with an amplitude of 15\,nWm$^{-2}$sr$^{-1}$ 
is in tension with our results. This could be due, to a large extent, to  our advantage in using high-$z$ blazar spectra.



{ The VHE spectra from blazars (especially FSRQs) are affected by internal absorption \citep{liu} due to multiple soft-radiation fields, mainly broad line region (BLR). However, considering different BLR parameters (size, thickness, temperature) \cite{mazintav} found no significant hardening in the IACT-accessible VHE range -- even in the extreme case of a UV radiation field (slope of -1/3), no $>$60\,GeV  spectral hardening is predicted. Since our high-redshift (z>0.25) subsamples are dominated by FSRQs, it is important to validate our results in the light of internal absorption. Our high redshift data contains 4 spectral points below  60\,GeV, hence we repeated the procedure by omitting these spectral points: we found no appreciable change in our results.}

{ It is worth emphasizing that several studies [eg: \citep{dea1, dea2, meyarhorns}] have suggested the possibility of $\gamma$-ray photon-ALP %
\footnote
{  Light ALP masses, $m_{a}\sim 10^{-10}-10^{-6}$\,eV, are relevant to the ALP-photon oscillation framework. On the other hand, 
heavy masses, $m_{a}\sim 1-10$\,eV, are relevant to the ALP-photon decay scenario (see section\,2, and later this section).}
oscillations in the presence of 
intergalactic magnetic fields. According to these models, the hardening or curvature of VHE spectra will be somewhat  
reduced by ALP oscillations.  So in principle these hypothetical oscillations may potentially explain the differences between the direct measurements and the baseline EBL.

The redshifted light from Pop-III stars is often associated with the additional EBL population invoked for the Optical-NIR band \citep{Kashlinsky2005}.  
The peak of the Pop-III stars emission can be explained using the Ly$\alpha$ photons generated by reprocessing the UV photons in the ionized gas 
around the stars \citep{Matsumoto, Fernandez}. Nevertheless, various models propose minute contribution (in which the peak of the spectra $P < 
0.5$\,nWm$^{-2}$sr$^{-1}$) from Pop-III stars [eg: { \cite{Sun, Yue, Fernandez, Inoue2013, silk}}]. 
On the other hand, { the models of \cite{Salvaterra}  reproduced direct NIR/EBL data peaking up to $\sim$ 70\,nWm$^{-2}$sr$^{-1}$ at $\sim$1-4\,$\mu$m}. 
Those authors assumed three different star formation efficiencies { ${\rm f}_{\ast}$ (defined as the fraction of baryons able to cool and form stars in a galaxy) for different assumptions of initial mass functions (IMF), by assuming a star formation cutoff $z_{\rm end} \sim 8.8$. They essentially reproduce similar EBL densities (${\rm f}_{\ast}$ = 0.53, 0.21, and 0.14 for Salpeter, heavy and very heavy IMFs respectively), which are all} clearly above the upper limit we set.

 In the framework of the \cite{Salvaterra} models, we modify the Pop-III spectra using lower ${\rm f}_{\ast}$. An acceptable population (w.r.t. VHE spectra)  is  
attained when the Pop-III contribution  is scaled down by a factor of $\sim$ 60, i.e., 1.6\% of the original star formation efficiency (${\rm f}_{\ast} \lesssim 0.008, 0.004, 0.002$ for Salpeter, heavy and very heavy IMFs respectively). The resulting Pop-III--derived photon spectrum is shown as red solid line in Fig.\,6\,(a). Such a small peak is in good agreement with 
our results, considering that { it is located} at $>$1\,$\mu$m. Moreover, the obtained Pop-III spectra, though not 
Gaussian shaped, is wider than the Gaussian spectral templates we assumed for the additional NIR photon fields: in this case the ULs shown in Fig.\,5 are even more stringent.  Arguably, then, a wider Pop-III contribution [eg: \cite{Helgason}] does not result in a significant difference in the measured flux versus estimates from galaxy counts.

In order to test a comparatively narrower Pop-III contribution together with our baseline EBL model, we adopted Pop-III 
spectra from \cite{Fernandez2010} which were derived using numerical simulations of reionization, { a value of} $f_{\ast}=0.01$, and 
an IMF by \cite{Larson}. 
 After applying  such composite EBL to the VHE spectra, we reject this extra population at a confidence level of 95\%. Scaling down the Pop-III spectrum by lower $f_{\ast}$ values we derive an UL, $f_{\ast}\lesssim 0.005$, to the star-formation efficiency for the \cite{Fernandez2010} model: the resulting population is shown as a blue solid line in Fig.\,6\,(a). We note that the peak of both (modified) Pop-III spectra in the figure are substantially lower than the directly measured fluxes.  A narrower Pop-III contribution (i.e., strong Ly-$\alpha$ contribution w.r.t to the Pop-III continuum) will explain a significant 
difference between the observed flux and the estimated flux through galaxy counts. A larger Ly$\alpha$ bump indicates the presence of significant 
number of massive Pop-III stars at relatively lower redshifts [z$\lesssim$8; \cite{Fernandez}].

 Employing the UL from \cite{mazinebl} and different models of zero metallicity (ZM) and low metallicity (LM) stars, \cite*{raue}  derived limits on the star formation rate (SFR), which is  $\sim$0.3-3$M_{\odot}$\,Mpc$^{-3}$\,yr$^{-1}$ in a redshift range of 7-14. In their study, they adopted a primordial stars EBL contribution of 5\,nWm$^{-2}$sr$^{-1}$ at $\sim$2 $\mu$m. Owing to a more restrictive upper limit ($\sim$1\,nWm$^{-2}$sr$^{-1}$) we can estimate a SFR $\sim$5 times lower than that of \cite{raue} for a similar stellar population. However, considering the wider nature of their stellar EBL model, our estimate can be improved if we employ the stellar EBL densities superposed on the baseline EBL and estimate ULs. In order to prove this, we use 3 samples each for ZM and LM stars using different SFR parameters: ascending slope $\alpha$=10, descending slope $\beta$=-2, and peak redshift z$_{\rm peak}$=7, 10, 13 (see, top panel of Fig.\,6 in \cite{raue} for details). We then estimate the SFR limits for this models, which turned out to be an order of magnitude lower than the original estimation. In Fig.\,7 we show the modified SFR limits as a green shaded region, while the original limits are shaded in yellow. 
  We also modify the SFR (corresponds to  $f_{\ast}= 0.005$; see also table\,3 of \cite{Fernandez2010}), which is shown as green solid line. Recent estimates of SFR from semianalytical models of the first stars \citep{hartwig} and cosmological simulations \citep*{sarmento,jaacks} are also shown in the figure.

 \begin{figure}
   \hspace{-0.5cm}
   \includegraphics[width=1.2\columnwidth]{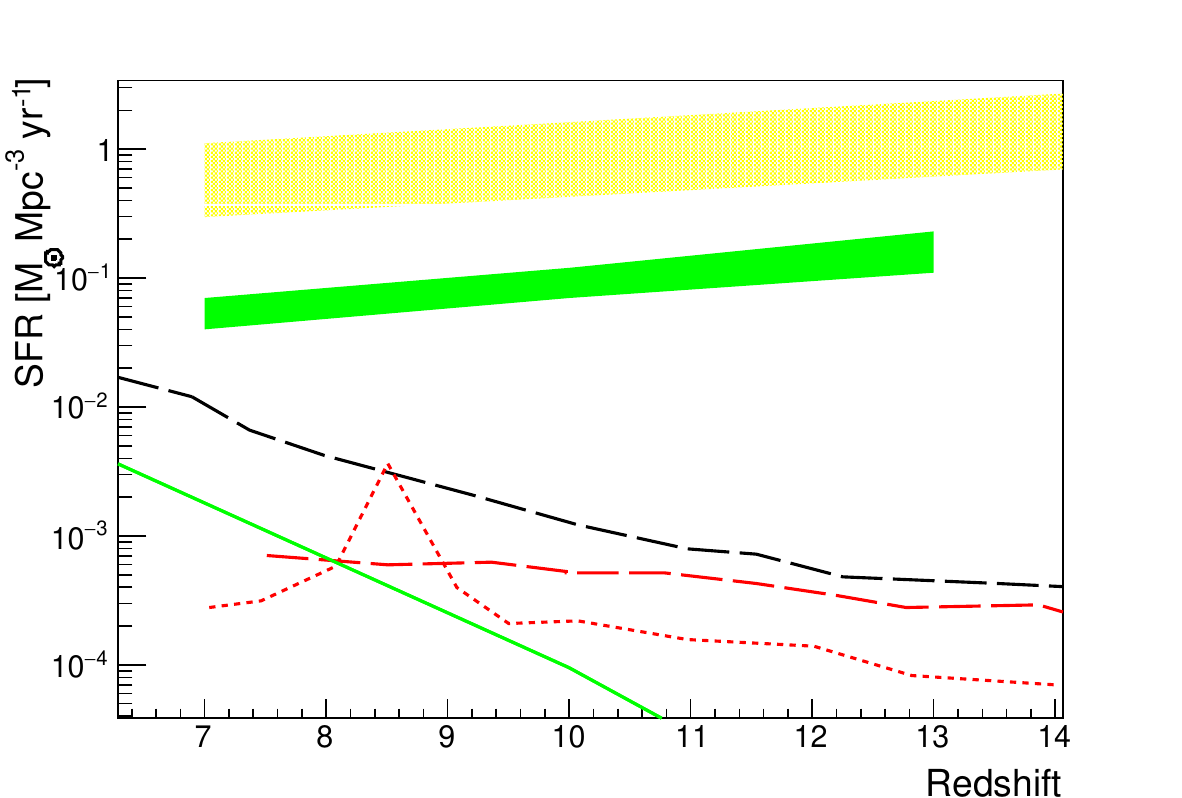}
   \vspace{-0.5cm}
   \caption{ Co-moving SFR of ZM and LM stars obtained using 6 limited parameter samples (see text for details) in \cite{raue} is shown as green shaded region, while the original estimation by \cite{raue} is shown in yellow. The green line indicates the modified SFR for the adopted Pop-III model in \cite{Fernandez2010}. Recent estimates from  cosmological simulations and semianalytical models are also shown- black dash line: Pop-III+ Pop-II \citep{hartwig}; red dash line: Pop-III \citep{sarmento}  red dotted line: Pop-III \citep {jaacks}. }
    \label{fig:example_figure}
\end{figure}

Axion (of mass $\rm{m_a}\sim$1-10\,eV)  decay into Optical-NIR photons is also a potential source of EBL excess above 0.5 $\mu$m. If single-mass 
axions are solely responsible for the significant excess [eg: \cite{Caputo}], { combining the wide range of permitted m$_a$ values and our UL} (in which no substantial emission above $\sim$ 1.0\,$\mu$m is allowed), we can estimate a lower limit on the axion 
mass of $\rm{m}_a \gtrsim 2.5$\,eV \citep{Gong}.  However, if we consider a  spectral distribution  for the axion mass, the resulting intensity spectrum is 
significantly wider [eg: \cite{Gong}] and fall outside our estimated excluded region. { To illustrate the case of ALP annihilation 
  contribution to the EBL we employed Equation (4) of \cite*{koro}. We have then repeated our analysis by varying the parameters of ALP mass ($m_a$) and two-photon coupling constant ($g_{a\gamma\gamma}$). As an example, see Fig.\,6\,(b) in which the red line corresponds to the combination of $m_{a}$=4.0eV and $g_{a\gamma\gamma}$=1.6$\times 10^{-10}$, while the blue line refers to $m_{a}$=2.8eV and $g_{a\gamma\gamma}$=1.8$\times 10^{-10}$. After testing various parameter combinations we have estimated an acceptable zone (see Fig.\,8) in the $m_a$-$g_{a\gamma\gamma}$ parameter space in which the  ALP contribution does not fail the t-test. The favoured region obtained by \cite{koro} contradicts our result because of the difference in basic assumptions of both approaches. They inferred no significant ALP contribution (hence, no extra bump, unlike it was shown in \cite{Neronov}) to the baseline EBL, while we look for the maximum allowed ALP contribution to the baseline EBL. The lower limit of our acceptable zone referes to the ALP contribution peak of 1\,nW\,m$^{-2}$sr$^{-1}$. We have also estimated a zone (red hatched region in Fig.\,8) with which the ALP contribution exceeds the limit of acceptable EBL at different wavelength bands. The $\rm{m_a}-g_{a\gamma\gamma}$ limits obtained under QCD framework (see \cite{Anastassopoulos} and references therein) is shown as gray band. The yellow band in the figure represents the limits obtained by the search of optical line emission from two-photon decay of axions \citep{grin}. }
 \begin{figure}
   \hspace{-0.5cm}
   \includegraphics[width=1.2\columnwidth]{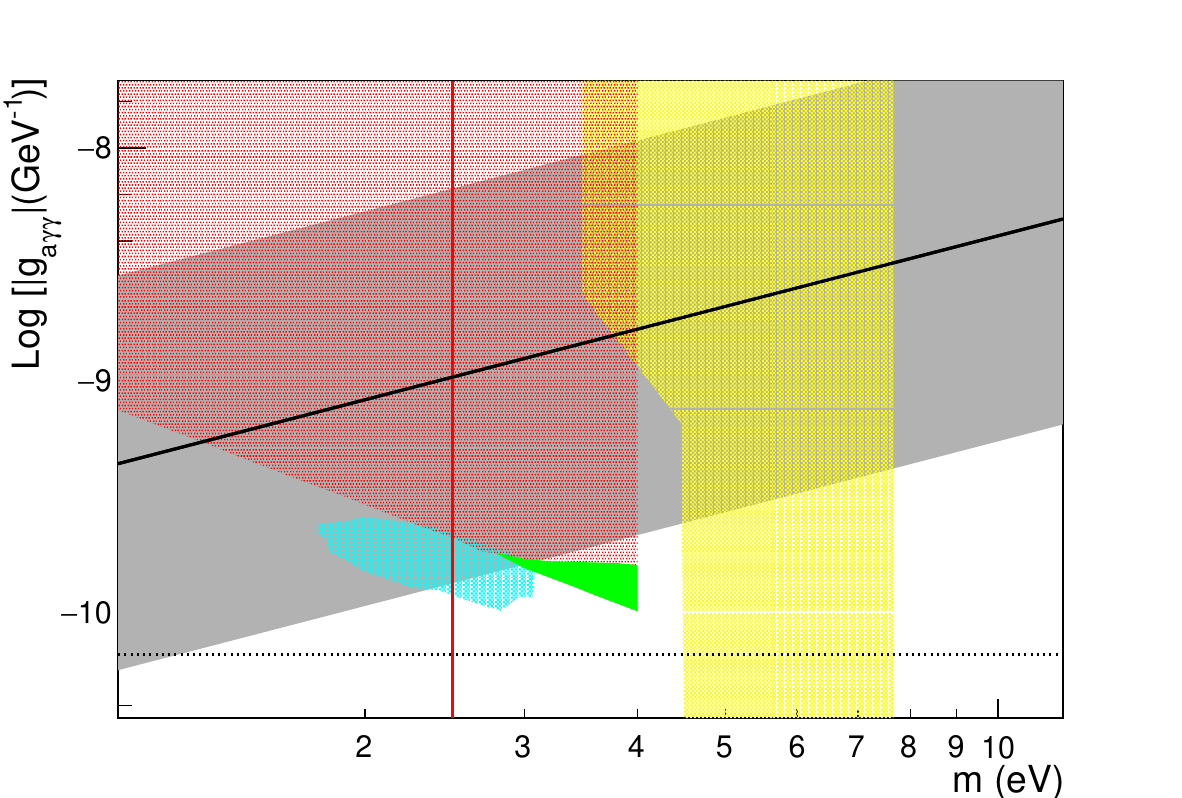}
   \vspace{-0.5cm}
   \caption{ The obtained ALP-two photon coupling constraints. The green triangle shows the favored region between the maximum allowed contribution and a minimum significant contribution (ie, $\sim$ 1nWm$^{-2}$sr$^{-1}$). The red hatched area is the region where a significant contribution to the EBL is excluded. The red vertical line corresponds to a mass $m_{a}=$2.5eV below which substantial ALP contribution to EBL is not favoured. The limits obtained from QCD models (gray shaded and black solid line), constraints obtained by observations of optical line emission from galaxy clusters (yellow hatched region), and non observation of excess energy loss in Horizontal Branch stars (dotted line) are also shown (see \cite{Anastassopoulos} and references therein, for details). The preferred region obtained by \cite{koro} is shown in Cyan.}
    \label{fig:example_figure}
\end{figure}

Contributions from IHL \citep{Zemcov} and 
faint galaxies \citep{Lauer2021} result in wider spectra. Therefore, they do not yield a substantial deviation between the observed flux and estimates from galaxy counts. On the other hand, additional populations from direct collapse of black holes and 
dark-matter--powered stars are not significant at $\lambda < 1\,\mu$m \citep{Yueb,Maurer,Kashlinsky2}.

\section{Conclusions}

There is a large observational and phenomenological uncertainty concerning the Optical-NIR band of the EBL. Employing 105 VHE spectra of 37 blazars 
(with known redshift) and a baseline EBL model, we estimated an exclusion region in which an additional population of EBL sources  may 
not be located. { We have then compared  it with recent direct EBL measurements and found that, at $\lambda \gtrsim  1\, \mu$m, the latter exceed our ULs. 
The main conclusions of this paper can be summarized as follows.

  1.  Above $\lambda \gtrsim 1\, \mu$m no significant photon population can be accommodated  on top of the baseline EBL model. This suggests the 
     possibility that the direct measurements at $\gtrsim$ 1\,$\mu$m may be affected by bad MilkyWay or zodiacal light subtractions.

  2. The recent New Horizon measurements \citep{Lauer2021} is well in agreement with our estimations. This may owe to the measurements being carried out at far distance from the Sun (beyond Pluto's orbit), where  zodiacal light is minimal.

  3. Given the proposed frequencies and/or spectral widths of their photon population, we can rule out any significant (i.e., amplitude $\gtrsim$ 1\,nW\,m$^{-2}$\,sr$^{-1}$) contributions from IHL, faint galaxies, direct  black-hole collapse, dark-matter--powered stars, and annihilation of a spectral distribution of axion mass \citep{Zemcov, Lauer2021,Yueb,Maurer,Kashlinsky2,Gong}.

  4. There is room for considerable photon population from Pop-III stars [\citep{Salvaterra,Fernandez2010} with modified star formation rates] and ALP ( $m_{a}>$\,2.5\,eV) annihilation \citep{koro}. 
   However, these components, if they exist, would not saturate any existing differences between the published direct measurements 
and the baseline EBL density.

  Since the high-$z$ blazar spectra are suitable to determine the forbidden region for the extra population, low-energy--threshold VHE 
observations of distant blazars  will be crucial in solving the ambiguity of Optical-NIR peak of the EBL. Direct 
measurements with space-borne instruments like JWST, CIBER-2, and upcoming  missions like Euclid-LIBRAE, SPHEREx and WFIRST will also be relevant \citep{ChengChang,Kashlinsky2021} at clarifying the issue.

\section*{Acknowledgements}

We acknowledge insightful comments and suggestions by the referee, Dr. Daniel Mazin, which have led to a substantial improvement of this paper.


\section*{Data Availability}


The data of this work will be shared upon request to the corresponding author.





\newcommand{\newblock}{}







\bsp	
\label{lastpage}
\end{document}